# Miniature Multi-Level Optical Memristive Switch Using Phase Change Material

Hanyu Zhang[1], Linjie Zhou[1,*], Liangjun Lu[1], Jian Xu[1], Ningning Wang[1], Hao Hu[1], B. M. A. Rahman[2], Zhiping Zhou[3], and Jianping Chen[1]

[1]State Key Laboratory of Advanced Optical Communication Systems and Networks, Shanghai Institute for Advanced Communication and Data Science, Department of Electronic Engineering, Shanghai Jiao Tong University, Shanghai 200240, China

[2]Department of Electrical and Electronic Engineering, City, University of London, London EC 1V 0HB, U.K.

[3]State Key Laboratory of Advanced Optical Communication Systems and Networks, Department of Electronics, Peking University, Beijing, 100871, China

ABSTRACT

The optical memristive switches are electrically activated optical switches that can memorize the current state. They can be used as optical latching switches in which the switching state is changed only by applying an electrical Write/Erase pulse and maintained without external power supply. We demonstrate an optical memristive switch based on a silicon MMI structure covered with nanoscale-size Ge$_2$Sb$_2$Te$_5$ (GST) material on top. The phase change of GST is triggered by resistive heating of the silicon layer beneath GST with an electrical pulse. Experimental results reveal that the optical transmissivity can be tuned in a controllable and repeatable manner with the maximum transmission contrast exceeding 20 dB. Partial crystallization of GST is obtained by controlling the width and amplitude of the electric pulses. Crucially, we also demonstrate that both Erase and Write operations, to and from any intermediate level, are possible with accurate control of the electrical pulses. Our work marks a significant step forward towards realizing photonic memristive switches without static power consumption, which are highly demanded in high-density large-scale integrated photonics.

INTRODUCTION

Photonic switches are core devices for energy- and cost-efficient photonic signaling and processing systems in cloud and data-intensive computing. Silicon photonics has emerged as a suitable platform for integrated optics owing to its low power consumption, small footprint, low cost by leveraging the advanced complementary metal-oxide-semiconductor (CMOS) manufacturing process [1, 2]. In the past few years, varieties of silicon photonic switches have been demonstrated on the silicon-on-insulator (SOI) platform with thermo-optic (TO) effect [3, 4], electro-optic (EO) effect [5-7], and micro-electro-mechanical-system (MEMS) actuation [8]. The MEMS actuator requires a high drive voltage of 40 V for turn-on and 25 V for turn-off [8]. The inefficient TO and EO effects provide a very small tuning range in the refractive index. Large-size Mach–Zehnder interferometers (MZIs) or narrow-band ring resonators are routinely used to realize optical switching. Moreover, the above three tuning methods are all volatile, which means that the switching state can only be maintained with an applied voltage. For TO and EO tuners, this results in a huge static power consumption.

On the other hand, memory switching effects have been studied in electronics for many years [9]. One important candidate for memory switching is based on the so-called resistive random-access memory (RRAM). Chalcogenide phase change materials (PCMs), such as germanium-antimony-tellurides (GST), have shown outstanding properties in RRAM due to their high Write and Read speeds, reversible phase transition, high degree of scalability, low power consumption, good data retention, and multi-level storage capability [10]. The reversible phase transitions between the amorphous and crystalline states could be triggered by thermal [11], optical [12] or electrical pulses [13] as long as the temperature of GST is raised above the crystallization or melting point. The phase change is non-volatile and therefore no power is required to keep the state. In addition to the electrical resistance contrast, a pronounced change in the optical property, including refractive index, extinction coefficient, and reflectance, also occurs in PCMs during the phase transition [14, 15]. Such features have led to the development of on-chip non-volatile photonic applications, such as nanopixel display [16, 17], photonic synapse [18, 19], optical computing [20], and optical switching [21-24].

Most interestingly, combining PCM with silicon photonics allows one to relate the worlds of resistive memory and optical switching, leading to the realization of a compact memristive switch. Optical memristive switches are digital optical switches with multiple different transmission states that can be stored for long periods without external voltage. Hence, it possesses characteristics from both an optical switch and an optical memory in one configuration [25, 26]. Optical memristive switches are particularly interesting for use as a new kind of optical field-programmable devices whose optical transfer functions can be electrically written and erased.

In this paper, we demonstrate an ultra-compact optical memristive switch based on a nanoscale GST-gated silicon multimode interferometer (MMI). The device exhibits a large transmission contrast when the GST changes between the amorphous and crystalline states. Both binary and multi-level optical switching has been achieved by controlling the crystallization degree of GST. We also investigate the dynamic processes during crystallization and amorphization of GST and their effect on the optical transmission.

DEVICE STRUCTURE

Figure 1(a) illustrates the structure of the optical memristive switch composed of an MMI crossing with a nanoscale GST on top. Input and output silicon waveguides are 500-nm wide and 220-nm high, connected to the MMI through two 4-μm-long linear tapers with a 1.2-μm tapered width. The MMI is 12-μm long and 2-μm wide. In the center of the MMI, another 1.7-μm-wide silicon strip is orthogonally crossed. Due to the self-imaging principle of MMI, light is focused into the center of the MMI and the crossed strip thus does not introduce significant loss. The crossed strip is heavily $P^{++}$-doped in the center with a doping strip width of 1 μm. At the end of the doping region, a 20/100-nm-thick Ti/Au layer is deposited on top to form ohmic contact. Therefore, the doping strip works as a resistive heater when an electrical pulse is applied through the gold (Au) electrodes. A small circular piece of 30-nm-thick GST layer with a diameter of $D_{GST}$ is placed in the center of the MMI. The GST layer is covered by another 30-nm-thick indium-tin-oxide (ITO) film to protect it from being oxidized when it is exposed in the air. Control of light wave propagation relies on the evanescent coupling between the MMI and the GST material. Change in extinction coefficient of the GST material after phase transition alters the optical transmittance through the MMI. As the GST is located at the light focusing spot, the interaction is maximized, favoring a high optical transmission tuning efficiency.

Finite-difference time-domain (FDTD) simulations were performed to calculate light propagation through the MMI crossing. Figure 1(b) illustrates the cross-sectional electrical-field intensity ($|E|^2$) distributions in the *x-y* and *y-z* planes when the GST is amorphous and crystalline. A significant reduction in optical transmission is observed when the GST changes from the amorphous to the crystalline state.

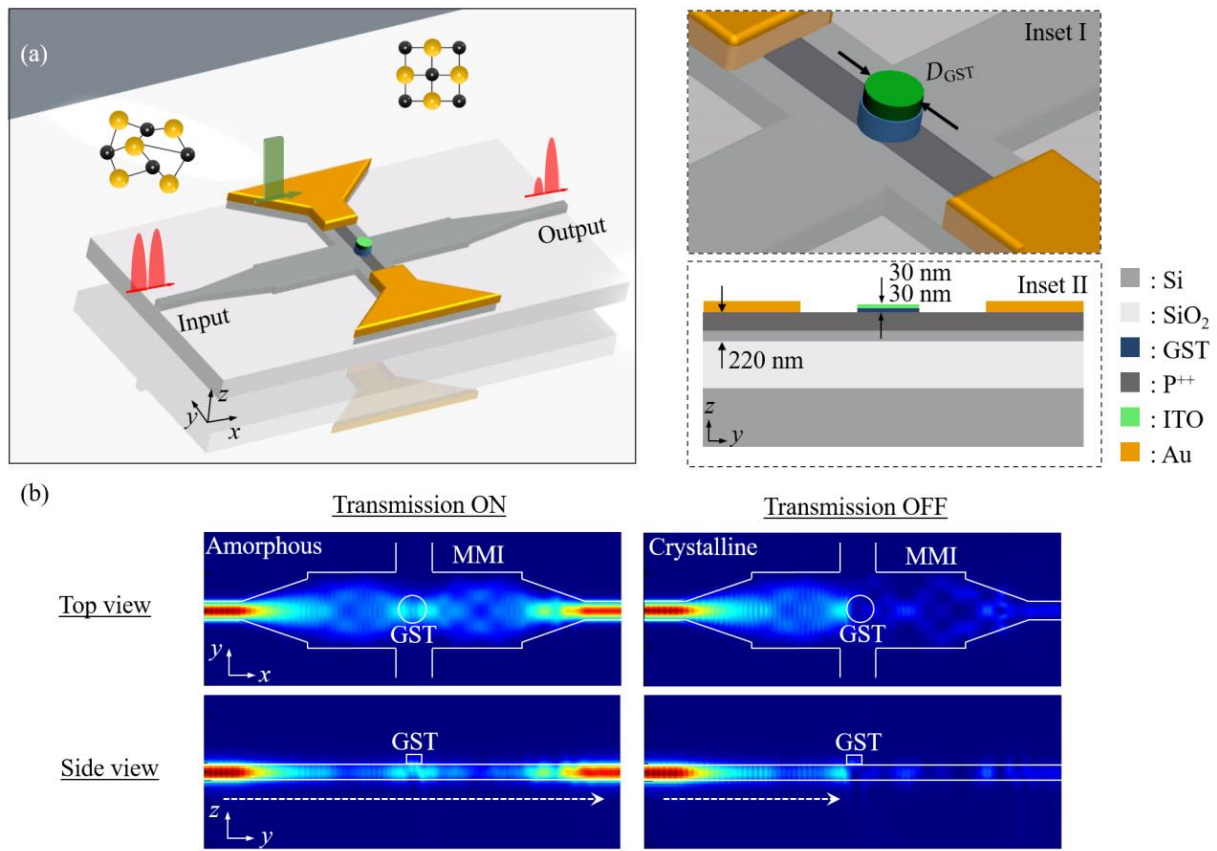

Figure 1. Device structure and optical wave propagation. (a) Schematic of the GST-gated silicon MMI switch. The inset I shows the zoom-in view of the active region. The inset II shows the transversal cross-sectional view of the active region. (b) Simulated electric-field intensity distributions in the MMI when the GST is in amorphous and crystalline states.

The non-volatile phase transition is induced by an electrical pulse (green arrow in Figure 1(a)). For amorphization, the GST is first melted and then quenched rapidly by applying a high-voltage electrical pulse (Write pulse) for a short time period. For crystallization, the GST is annealed by applying a medium-voltage electrical pulse (Erase pulse) at a temperature between the crystallization and melting points for a time long enough to recover the atomic order of GST. For optical memristive switch applications, the nonvolatility is beneficial since no static power is required to keep the current state and only a short low-power pulse is required to change the phase of the nanoscale GST material.

The devices were fabricated (see Supplementary Material S1) using an SOI wafer with a 220-nm-thick silicon layer on top of a 2-μm-thick buried oxide layer. Figure 2(b) presents the false-colored scanning electron microscope (SEM) image of the coupling region, where the GST/ITO patch is clearly discerned.

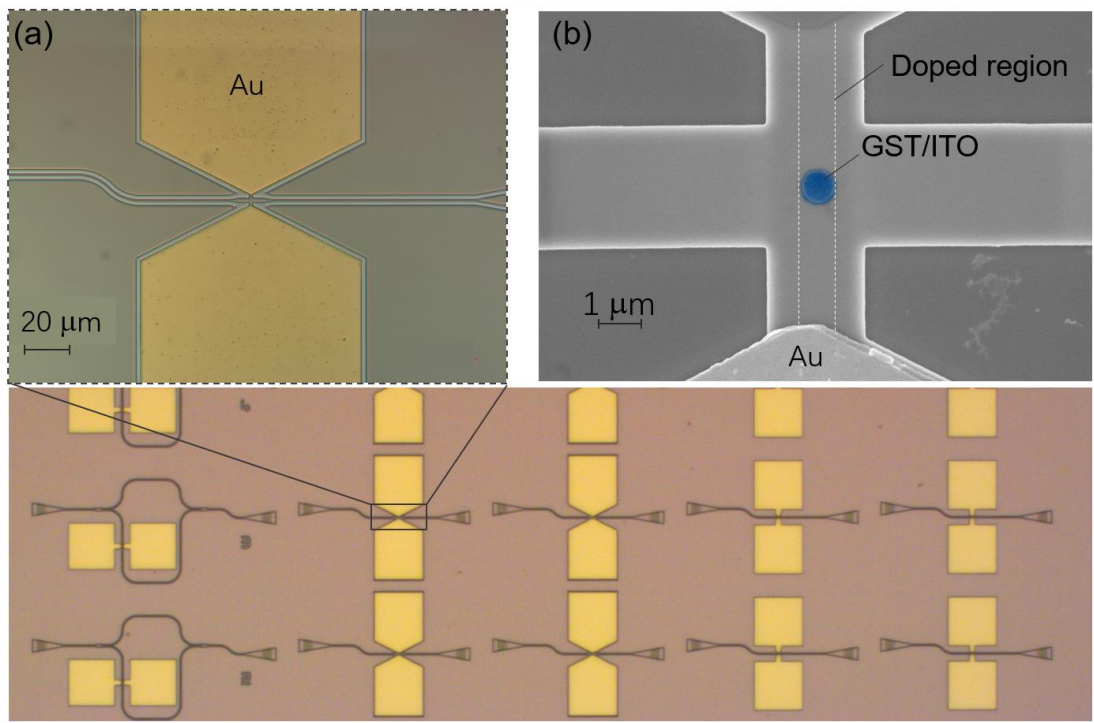

Figure 2. Fabricated devices. (a) Optical microscope image of the fabricated device. (b) False-colored SEM image of the MMI crossing with GST/ITO in the center.

EXPERIMENTAL RESULTS

CRYSTALLIZATION PROCESS. The GST crystallization (Erase) process determines the optical transmission dynamics during switch-off. The optical transmission ($T_0$) when GST is in the full crystalline state is defined as the baseline. The normalized transmission change is then defined as $\Delta T/T_0 = (T-T_0)/T_0$. We first investigated the dynamic change in transmission with a 1-μm-diameter GST when we applied to the device a rectangular Erase pulse with a fixed duration of 100 ns and a varying amplitude from 3.5 V to 6 V, as shown in Figure 3(a). The GST was initially at the amorphous state. When the Erase pulse amplitude was set to be low enough, the GST phase transition was inhibited. In the period of the Erase pulse, the optical transmission continuously decreases. At the end of the pulse, the transmission rises up until it recovers its original value. The recovery time is about 100 ns, determined by the heat out-diffusion process during cooling down. However, as the pulse amplitude increases to exceed 4.5 V, GST begins to be partially crystallized and the transmission continuously drops after the excitation pulse. When the material is fully crystallized, the transmission goes down to the minimum and remains unchanged with the further increasing pulse amplitude. Figure 3(b) shows the maximum transmission change as a function of pulse amplitude. It can be seen that once the temperature exceeds the crystallization threshold, the amount of transmission change increases rapidly until finally saturates at the full crystalline state.

We next fixed the Erase pulse amplitude at 3.5 V while the pulse duration increases from 40 ns to 200 ns. Figure 3(c) shows the measured dynamic responses. The transmission returns to the initial level when the pulse duration is shorter than 140 ns. It begins to crystalline for pulse duration larger than 140 ns and is fully crystallized when the pulse duration is longer than 180 ns. Figure 3(d) shows the maximum transmission change as a function of pulse duration for various pulse amplitudes. The transmission is relatively insensitive for low-amplitude pulses. The threshold duration to initiate crystallization becomes shorter for high-amplitude pulses. It can also be seen from the experimental results that besides the crystalline and amorphous states, there also exist multiple intermediate states with a mixture of crystalline and amorphous GST. Utilizing these intermediate states can further increase the waveguide tuning freedom.

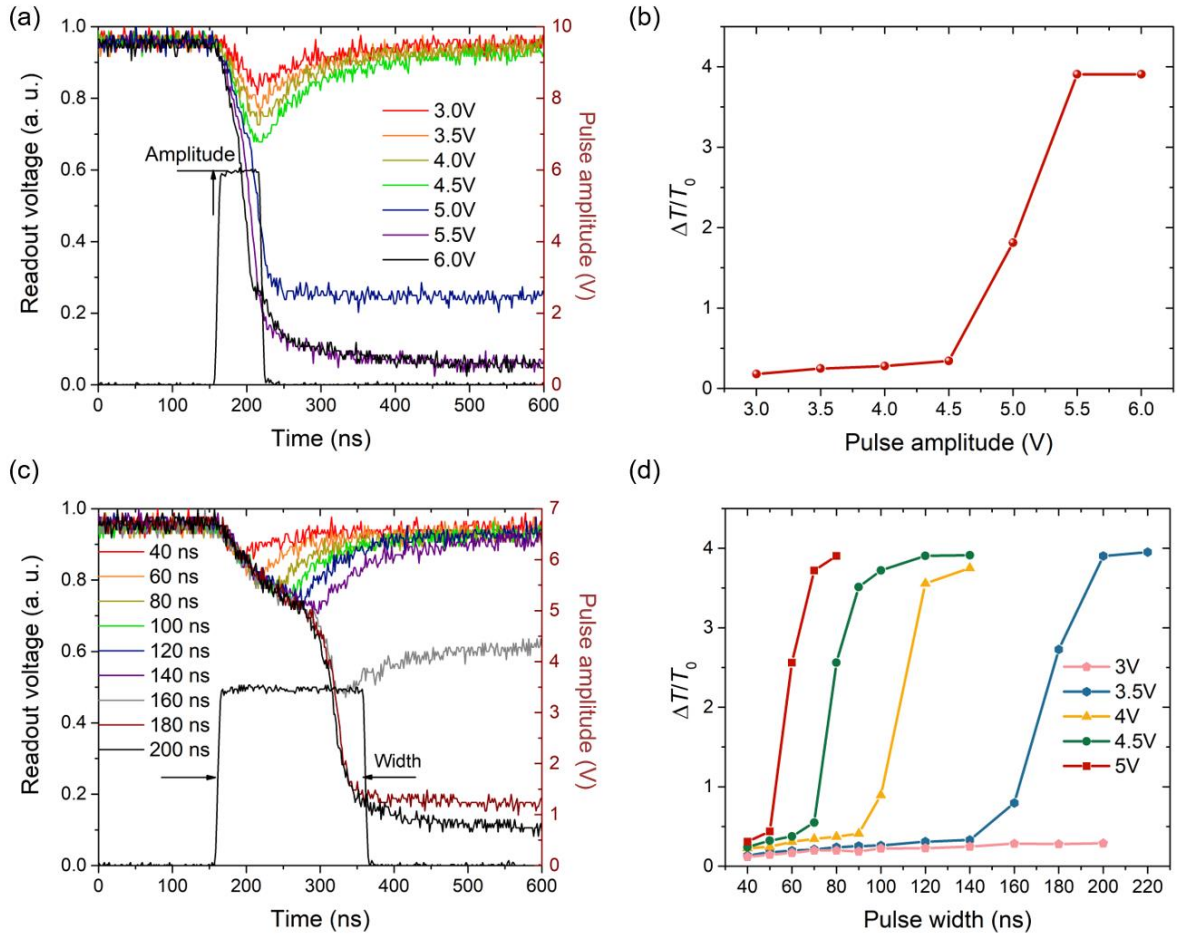

Figure 3. Dynamic response of optical transmission during the Erase process. (a) The input electrical pulse has a fixed duration and a varying amplitude. (b) The maximum transmission change varies with the pulse amplitude. (c) The input electrical pulse has a fixed amplitude and a varying duration. (d) The maximum transmission change varies with the pulse duration and amplitude.

AMORPHIZATION PROCESS. The GST amorphization (Write) process determines the optical transmission dynamics during switch-on. In the amorphization process, the GST needs to be heated up to the melting temperature (650°C) and then quenched rapidly by applying a high-voltage electrical Write pulse for a short time period. Figure 4(a) shows the temporal response when the pulse amplitude is varied and the pulse duration is fixed at 20 ns. The GST is initially at the fully crystalline state. When the pulse amplitude is low (<10.8 V) so that the GST is not heated to the melting point, the optical transmission first increases and then drops to the initial level. When the melting temperature is reached with a higher pulse amplitude (11V~11.6V), the GST is partially amorphized and the degree of amorphization increases with the pulse amplitude until it is fully amorphized (11.8 V). However, when the pulse amplitude is further increased (12 V), GST may be ablated with an increased loss, deteriorating the device performance. Figure 4(b) shows the maximum transmission change as a function of the pulse amplitude.

Figure 4(c) shows the amorphization dynamic process when the Write pulse has a fixed amplitude but a varying duration from 10 ns to 20 ns. The transmission rises and drops to the initial low level when the pulse duration is shorter than 16 ns. After that, the amorphization begins and the transmission level goes up rapidly. Figure 4(d) illustrates the maximum transmission change varying with the pulse duration.

From Figures 4 (a) and (c), one sees that the response time of the amorphization process is around 100 ns, governed by the gradual increasing transmission after the excitation pulse. It is believed to be related to the heat dissipation process when the GST is quenched.

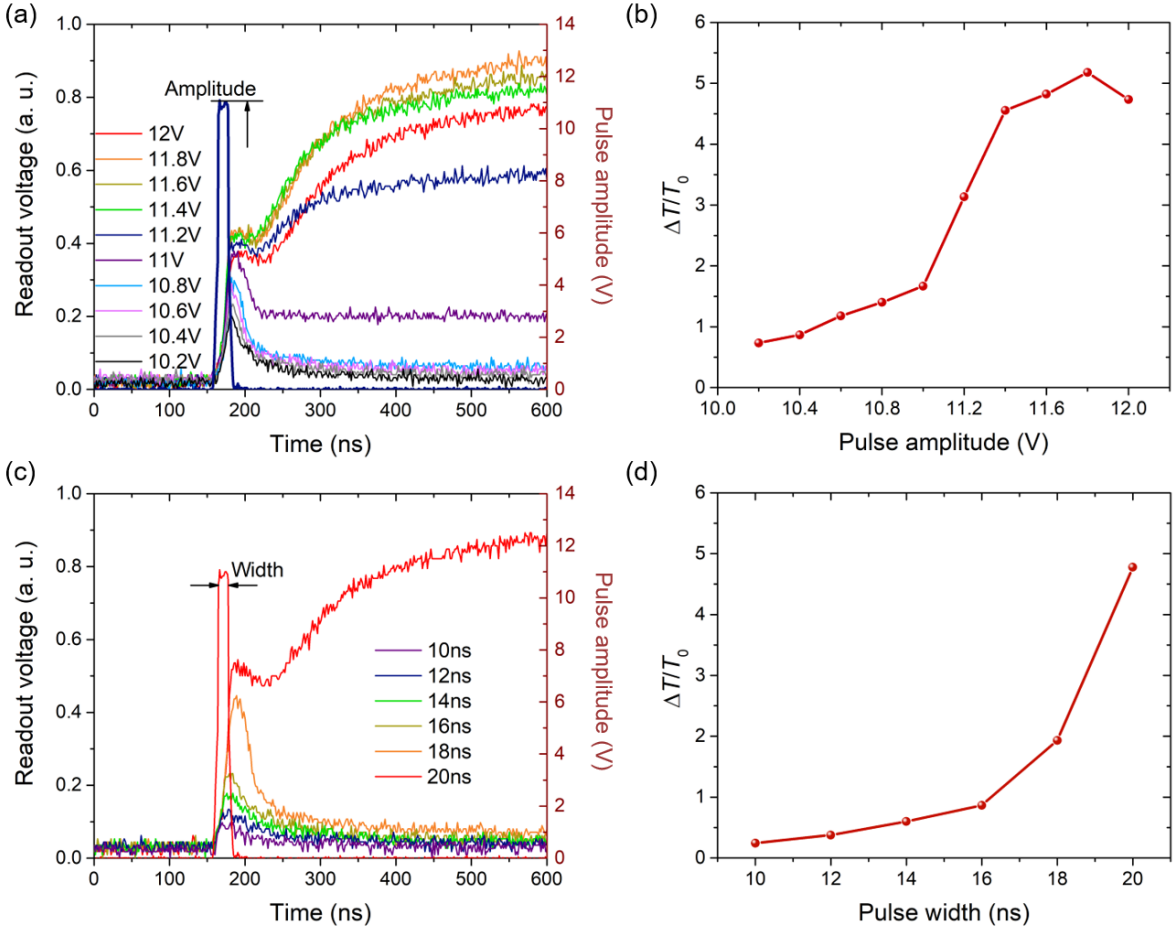

Figure 4. Dynamic response of optical transmission during the Write process. (a) The input electrical pulse has a fixed duration and a varying amplitude. (b) The maximum transmission change varies with the pulse amplitude. (c) The input pulse has a fixed amplitude and a varying duration. (d) The maximum transmission change varies with the pulse duration.

TRANSMISSION CONTRAST AND OPERATION SPEED. We next investigated the effects of different sizes of GST on the transmission contrast and the operating speed. We fabricated three devices with the GST diameter being 0.75 μm, 1 μm, and 1.2 μm. The results are summarized in Table 1. The silicon doping region has a fixed width of 1 μm for all devices. We used the same Write and Erase pulses to initiate the phase transition for all devices. The Supplementary Material S3 shows an example of transmission spectra for two phase-change cycles. The measured result suggests that the device can operate over a bandwidth of more than 100 nm

(limited by measurement). For the device with a 1.2-μm-diameter GST, the insertion loss increases to 4 dB and the transmission contrast is improved to 18.59 dB. It reveals that a higher transmission contrast is obtained from the device with a larger-size GST. Because the doping region is narrower than the GST area, the GST is not uniformly heated up. Therefore, the GST is not completely amorphized after a single Write pulse, leading to a higher insertion loss due to the residual crystalline GST.

Table 1. Comparison of optical memristive switches with three difference sizes of GST

| Diameter of GST film, $D_{GST}$ (μm) | Insertion loss (dB) | Transmission contrast (dB) |
| --- | --- | --- |
| 0.75 | 0.62±0.044 | 2.92±0.021 |
| 1 | 0.98±0.048 | 10.39±0.045 |
| 1.2 | 4.03±0.019 | 18.59±0.15 |

It reveals that the transmission contrast exceeding 20 dB are possible with a larger-size GST at the cost of a higher loss. The insertion loss can be decreased by slightly widening the doping region to incorporate the entire GST material. On the other hand, the recently discovered new optical phase change material $Ge_2Sb_2Se_4Te_1$ (GSST) exhibits significantly reduced extinction coefficient compared to GST in the telecom wavelength range [27]. Therefore, it is possible to realize a memristive switch based on GSST with an even lower insertion loss. In the Supplementary Material S4 shows the transmission dynamic responses to the Write and Erase pulses for three devices with different GST sizes. It reveals that the response time of the writing process is longer and increases with the GST size. There is a fundamental tradeoff between the transmission contrast and the operation speed.

In our device, the doping region is a rectangular strip beneath the GST, and therefore, heat is uniformly generated along the thin strip. Given that the GST is only a micrometer-size circular disk, a considerable heat is thus wasted. The doping window can be further optimized to have a reduced waist in the GST region so that heat can be more concentrated in the center. The gap size between the two Au electrodes is another key parameter affecting the electrical power required for the Write/Erase process. A narrower gap gives a smaller silicon resistance, which helps to reduce the pulse voltage and improve the power efficiency. The operation speed is also increased with the reduced heat dissipation. It is expected that switching time of less than one hundred nanosecond is achievable.

OPTICAL MEMRISTIVE SWITCHING. Figure 5(a) shows the optical waveform when the transmission changes between the ON-state (amorphous) and the OFF-state (crystalline). A 20-ns-wide pulse of 11.2 V amplitude was used for re-amorphization and a 100-ns-wide pulse of 4.5 V amplitude was used for crystallization. The resistance of the $P^{++}$-doped region is approximately 250 Ω. The energy consumption was calculated to be 10 nJ for the re-amorphization process and 8.2 nJ for the crystallization process. We also measured the recyclability of our optical memristive switch. Figure 5(b) shows the experimental results for repetitive switching over 50 cycles. Since a hundred thousand times of rewriting has been achieved with phase-change materials, the number of switching cycles in our device can be further improved [28]. The electrically-driven GST memristive switch can not only realize binary switching, but is also capable of multi-level operation. This multi-level operation relies on the intermediate states with a mixture of crystalline and amorphous GST. The number of the intermediate states can be obtained by controlling the degree of crystallization in a single device. Figure 5(b) shows the switching operation with five clear distinct levels. The initial phase of GST is fully crystalline with a low transmission (level 0).

It returns to the high level (level 4) after amorphization at time $t_1$ using a single Write pulse with 20 ns duration and 520 mW peak power (11.4V pulse amplitude) to ensure that all GST is re-amorphized. The Write operation (return to Level 4) can be performed at any intermediate level, such as from Level 1 and 2 to Level 4 at time $t_5$ and $t_{12}$, respectively. The Erase operation (return to Level 0) is also possible from any intermediate state, such as from Level 2, 3, 4 to Level 0 at $t_{15}$, $t_8$, $t_2$, respectively. The Erase pulse has a duration of 100 ns and a peak power of 90 mW (4.75 V pulse amplitude). The partial crystallization of GST (corresponding to Level $n$) can be obtained from Level 4 by using a group of weak Erase pulses with 100 ms period, 140 ns duration, and 50 mW power (3.5 V pulse amplitude). The number of pulses required for partial crystallization is 5-$n$. For example, four, three, and two weak Erase pulses are used to switch from Level 4 to Level 1, 2, 3 during the time periods of $t_4$-$t_5$, $t_{10}$-$t_{11}$, $t_6$-$t_7$, respectively. This represents a crucial advancement, because it allows one to arbitrarily switch the transmission to the desired level using a combination of Write and Erase pulses without pre-knowledge of the current level. Each weak Erase pulse has the same duration and amplitude. As a result, the first weak Erase pulse does not have a significant effect on the crystallization process. In fact, a sequence of consecutive pulses with gradually decreasing amplitude can be used to generate more evenly distributed intermediate levels [12, 29].

We use different colors to code the bit levels, the result shows that five clearly distinguishable levels can be obtained. In fact, more transmission levels can be programmed, limited only by the noise level of the photodetector. These exciting results of our memristive optical switch demonstrate that any intermediate level can be reached just by applying the appropriate Write or Erase pulses. Such capabilities provide a huge leap forward in terms of functionality and will be crucial for applications in optical latching switch, optical computing, and reconfigurable photonic circuits.

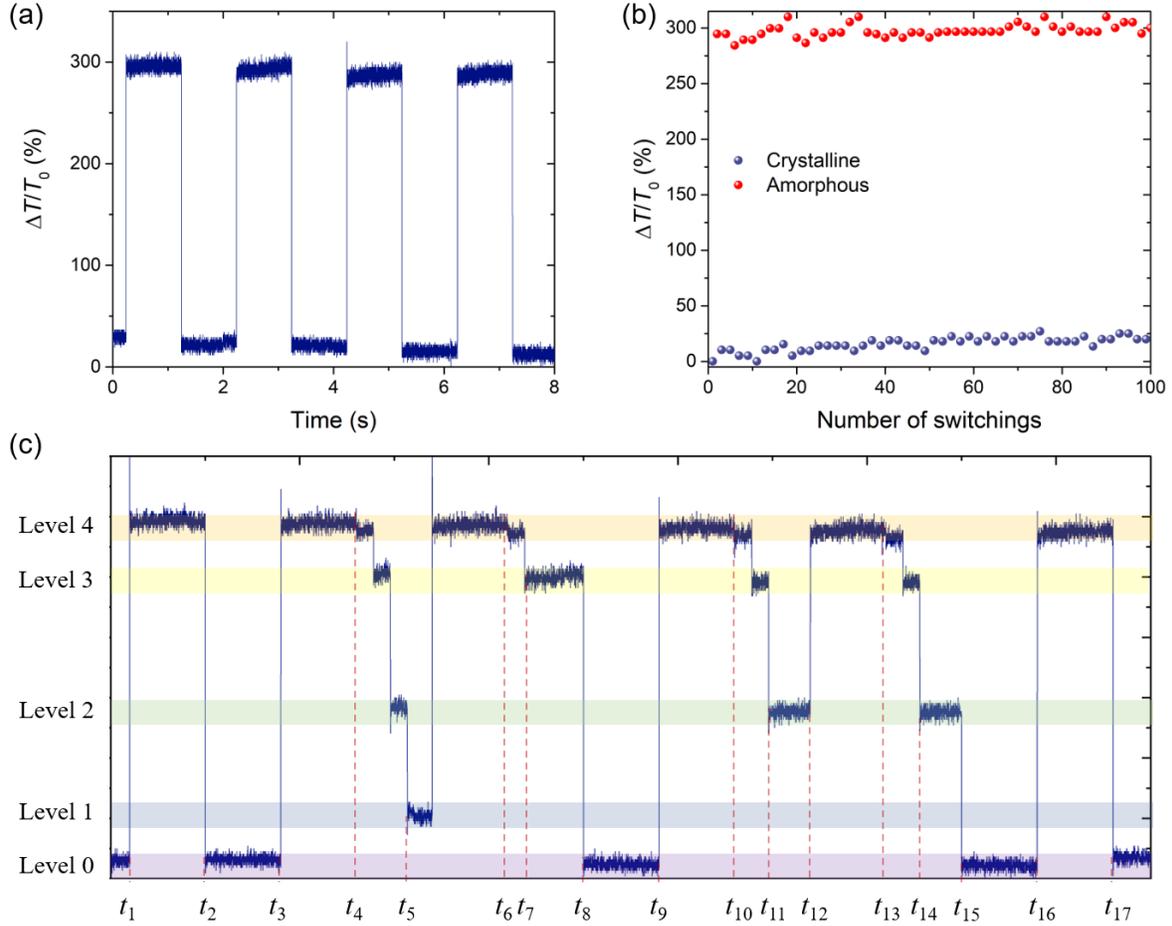

Figure 5. (a) Binary-level temporal response when the device is switched between the ON-state (amorphous) and the OFF-state (crystalline). (b) Transmission contrast over more than 50 switching cycles. (c) Multi-level switching of device with each middle level corresponding to a partially-crystalline state.

CONCLUSION

We have demonstrated a memristive optical switch whose optical transmission level can be electrically written and erased by combining a silicon MMI and the phase change material GST. The switching state change is based on the electrical pulse induced phase transition of a µm-size

GST material placed on top of the MMI. The measurement result reveals that the transmission contrast exceeding 20 dB are possible over the wavelength range of 1,500 to 1,600 nm. We have investigated the transient optical responses of our memristive switch in response to electrical pulses with varying amplitude and duration. This study helps us to better control the Erase/Write pulse, so as to accurately control the degree of crystallization/re-amorphization of GST. Both binary-level and multi-level switching operations are thus possible. Our method provides a new tool for an on-chip programmable photonic circuit with the non-volatile operation. This successful implementation of an optical memristive switch actuated by GST marks a significant step forward in realizing an ultra-small and low-power consumption optical circuit and set the stage for further exciting new developments in phase-change enabled silicon photonics.

## ASSOCIATED CONTENT

**Supplementary material (PDF)**

S1. Fabrication process flow

S2. Experimental setup

S3. Transmission contrast

S4. GST size effect on operation speed

## AUTHOR INFORMATION

**Corresponding Author**

E-mail: ljzhou@sjtu.edu.cn

**Author Contributions**

H. Zhang designed, fabricated and characterized the devices, and drafted the manuscript. L. Zhou and L. Lu initiated and led the project. J. Xu contributed to device fabrication. N. Wang and H. Hu aided in the experiment. B. M. A. Rahman, J. Chen and J. Zhou supervised the research. All authors contributed to technical discussions and writing the paper.

**Notes**

The all authors declare no conflict of interest.

**Funding Sources**

National Natural Science Foundation of China (NSFC) (61535006, 61705129), Shanghai Municipal Science and Technology Major Project (2017SHZDZX03).

ACKNOWLEDGMENT

The authors would like to thank the support from National Natural Science Foundation of China (NSFC) (61535006, 61705129), Shanghai Municipal Science and Technology Major Project (2017SHZDZX03). The authors also thank the Center for Advanced Electronic Materials and Devices (AEMD) of Shanghai Jiao Tong University (SJTU) for support in device fabrication.